\documentstyle[sprocl]{article}

\def\be{\begin{equation}}
\def\ee{\end{equation}}
\def\bea{\begin{eqnarray}}
\def\eea{\end{eqnarray}}
\def\lesssim{\mathrel{\hbox{\rlap{\hbox{\lower4pt\hbox{$\sim$}}}\hbox{$<$}}}}
\def\gtrsim{\mathrel{\hbox{\rlap{\hbox{\lower4pt\hbox{$\sim$}}}\hbox{$>$}}}}

\begin{document}

%%% start CERN preprint title page %%%%%%%%%%%%%

\begin{titlepage}
 
\begin{flushright}
CERN-TH/98-288\\
hep-ph/9809216
\end{flushright}
 
\vspace{1.5cm}
 
\begin{center}
\Large\bf CP Violation in B Decays and Strategies for\\
\vspace{0.3truecm}
Extracting CKM Phases
\end{center}
 
\vspace{1.2cm}
 
\begin{center}
Robert Fleischer\\
{\sl Theory Division, CERN, CH-1211 Geneva 23, Switzerland}
\end{center}
 
\vspace{1.3cm}

\begin{center}
{\bf Abstract}\\[0.3cm]
\parbox{11cm}{
A brief review of CP violation in the $B$-meson system and of strategies 
to determine the angles $\alpha$, $\beta$ and $\gamma$ of the unitarity 
triangle of the CKM matrix is given. Both general aspects and some
recent developments are discussed, including a critical look at the
``benchmark'' modes, CP violation in penguin decays, $B_s$ decays in the 
light of the width difference $\Delta\Gamma_s$, charged $B$ decays, and 
strategies to probe the CKM angle $\gamma$ with $B\to\pi K$ modes.}
\end{center}
 
\vspace{1.5cm}
 
\begin{center}
{\sl Invited talk given at the\\
4th International Workshop on Particle Physics Phenomenology,\\
Kaohsiung, Taiwan, 18--21 June 1998\\
To appear in the Proceedings}
\end{center}
 
\vspace{1.5cm}
 
\vfil
\noindent
CERN-TH/98-288\\
September 1998
 
\end{titlepage}
 
\thispagestyle{empty}
\vbox{}
\newpage
 
\setcounter{page}{1}
 
%%% end CERN preprint title page %%%%%%%%%%%%%

\title{CP VIOLATION IN B DECAYS AND STRATEGIES FOR EXTRACTING CKM PHASES}

\author{ ROBERT FLEISCHER }
 
\address{Theory Division, CERN\\ 
CH-1211 Geneva 23, Switzerland\\
E-mail: Robert.Fleischer@cern.ch}


\maketitle\abstracts{
A brief review of CP violation in the $B$-meson system and of strategies 
to determine the angles $\alpha$, $\beta$ and $\gamma$ of the unitarity 
triangle of the CKM matrix is given. Both general aspects and some
recent developments are discussed, including a critical look at the
``benchmark'' modes, CP violation in penguin decays, $B_s$ decays in the 
light of the width difference $\Delta\Gamma_s$, charged $B$ decays, and 
strategies to probe the CKM angle $\gamma$ with $B\to\pi K$ modes.}

\section{Setting the Scene}\label{intro}
At present, CP violation is one of the least experimentally explored 
phenomena of the Standard Model, and is very promising in the search for
indications of ``new physics'' at future experiments. In order to 
accomplish this task, it is crucial to have CP-violating processes 
available that can be analysed in a reliable way within the framework 
of the Standard Model, where CP violation is closely related to the 
Cabibbo--Kobayashi--Maskawa matrix (CKM matrix),\,\cite{ckm} connecting 
the electroweak eigenstates of the $d$, $s$ and $b$ quarks with their mass 
eigenstates. As far as CP violation is concerned, the central feature is 
that -- in addition to three generalized Cabibbo-type angles -- also 
a {\it complex phase} is needed in the three-generation case to parametrize 
the CKM matrix. This complex phase is the origin of CP violation within 
the Standard Model.
  
A closer look shows that CP-violating observables are proportional 
to the following combination of CKM matrix elements:\,\cite{jarlskog}
\begin{equation}
J_{\rm CP}=\pm\,\mbox{Im}\left(V_{i\alpha}V_{j\beta}V_{i\beta}^\ast 
V_{j\alpha}^\ast\right)\quad(i\not=j,\,\alpha\not=\beta)\,,
\end{equation}
representing a measure of the ``strength'' of CP violation in the 
Standard Model. Since $J_{\rm CP}={\cal O}(10^{-5})$, CP 
violation is a small effect. In many scenarios of new 
physics,\,\cite{new-phys} several additional complex couplings 
are present, leading to new sources of CP violation.

Concerning phenomenological applications, the parametrization 
\begin{equation}\label{wolf2}
\hat V_{\mbox{{\scriptsize CKM}}} =\left(\begin{array}{ccc}
1-\frac{1}{2}\lambda^2 & \lambda & A\lambda^3 R_b\, e^{-i\gamma} \\
-\lambda & 1-\frac{1}{2}\lambda^2 & A\lambda^2\\
A\lambda^3R_t\,e^{-i\beta} & -A\lambda^2 & 1
\end{array}\right)+\,{\cal O}(\lambda^4)
\end{equation}
with $\lambda=0.22$, $A\equiv|V_{cb}|/\lambda^2=0.81\pm0.06$, 
$R_b\equiv|V_{ub}/(\lambda V_{cb})|=\sqrt{\rho^2+\eta^2}=0.36\pm0.08$, and 
$R_t\equiv|V_{td}/(\lambda V_{cb})|=\sqrt{(1-\rho)^2+\eta^2}={\cal O}(1)$
turns out to be very useful. It is a modification of the Wolfenstein 
parametrization,\,\cite{wolf} exhibiting not only the hierarchy 
of the CKM elements, but also the dependence on the angles 
$\beta=\beta(\rho,\eta)$ and $\gamma=\gamma(\rho,\eta)$ of the 
usual ``non-squashed'' unitarity triangle of the CKM matrix.\,\cite{ut}

Although the discovery of CP violation\,\cite{ccft} goes back to 1964,
so far this phenomenon has been observed only within the neutral $K$-meson 
system, where it is described by two complex quantities, called $\varepsilon$ 
and $\varepsilon'$, which are defined by the following ratios of decay 
amplitudes:
\begin{equation}\label{defs-eps}
\frac{A(K_{\rm L}\to\pi^+\pi^-)}{A(K_{\rm S}
\to\pi^+\pi^-)}=\varepsilon+\varepsilon',\quad
\frac{A(K_{\rm L}\to\pi^0\pi^0)}{A(K_{\rm S}
\to\pi^0\pi^0)}=\varepsilon-2\,\varepsilon'.
\end{equation}
While $\varepsilon=(2.280\pm0.013)\times e^{i\frac{\pi}{4}}\times 10^{-3}$
parametrizes ``indirect'' CP violation, originating from the fact that
the mass eigenstates of the neutral kaon system are not CP eigenstates,
the quantity Re$(\varepsilon'/\varepsilon)$ measures ``direct'' CP violation 
in $K\to\pi\pi$ transitions. The CP-violating observable $\varepsilon$ plays 
an important role to constrain the unitarity triangle\,\cite{bf-rev} and 
implies -- using reasonable assumptions about certain hadronic parameters --
in particular a positive value of the Wolfenstein parameter~$\eta$. Despite 
enormous experimental efforts, the question of whether 
Re$(\varepsilon'/\varepsilon)\not=0$ could not yet be answered. However, 
in the near future, this issue should be clarified by improved measurements 
at CERN and Fermilab, as well as by the KLOE experiment at DA$\Phi$NE. 
Unfortunately, the calculations of Re$(\varepsilon'/\varepsilon)$ are very 
involved and suffer at present from large hadronic 
uncertainties.\,\cite{bf-rev}\, Consequently, this observable will not allow 
a powerful test of the CP-violating sector of the Standard Model, unless 
the hadronic matrix elements of the relevant operators can be brought under 
better control. Probably the major goal of a possible future observation 
of Re$(\varepsilon'/\varepsilon)\not=0$ would hence be the unambiguous 
exclusion of ``superweak'' models of CP violation.\,\cite{superweak}

In order to test the Standard Model description of CP violation, the 
rare decays $K_{\rm L}\to\pi^0\nu\overline{\nu}$ and 
$K^+\to\pi^+\nu\overline{\nu}$ are more promising and may allow a 
determination of $\sin(2\beta)$ with respectable accuracy.\,\cite{bb}\,
Yet the kaon system by itself cannot provide the whole picture of CP 
violation. Consequently, it is essential to study CP violation outside this 
system. In this respect, the $B$-meson system appears to be most promising, 
which is also reflected by the tremendous experimental efforts at the future 
$B$-factories. There are of course also other interesting systems to 
explore CP violation and to search for signals of new physics, for instance 
the $D$-meson system, where sizeable mixing or CP-violating effects would 
signal new physics because of the tiny Standard Model ``background''. In 
the following, we shall focus on $B$ decays.

\section{The Central Target: CP Violation in the B System}\label{CP-B}
With respect to testing the Standard Model description of CP violation, 
the major role is played by non-leptonic $B$ decays, which can be divided 
into three decay classes: decays receiving both tree and penguin 
contributions, pure tree decays, and pure penguin decays. There are two 
types of penguin topologies: gluonic (QCD) and electroweak (EW) penguins 
related to strong and electroweak interactions, respectively. Because of 
the large top-quark mass, also the latter operators play an important role 
in several processes.\,\cite{rev}

To analyse non-leptonic $B$ decays theoretically, one uses low-energy 
effective Hamiltonians, which are calculated by making use of the operator 
product expansion, yielding transition matrix elements of the following 
structure:
\begin{equation}\label{ee2}
\langle f|{\cal H}_{\rm eff}|i\rangle\propto\sum_k C_k(\mu)
\langle f|Q_k(\mu)|i\rangle\,.
\end{equation}
The operator product expansion allows us to separate the short-distance
contributions to this transition amplitude from the long-distance 
ones, which are described by perturbative Wilson coefficient 
functions $C_k(\mu)$ and non-perturbative hadronic matrix elements 
$\langle f|Q_k(\mu)|i\rangle$, respectively. As usual, $\mu$ denotes an 
appropriate renormalization scale. 

In the case of $|\Delta B|=1$, $\Delta C=\Delta U=0$ transitions, which 
will be of particular interest for the following discussion, we have
\begin{equation}\label{e3}
{\cal H}_{\mbox{{\scriptsize eff}}}={\cal H}_{\mbox{{\scriptsize 
eff}}}(\Delta B=-1)+{\cal H}_{\mbox{{\scriptsize eff}}}(\Delta B=-1)^\dagger,
\end{equation}
where 
\begin{equation}\label{e4}
{\cal H}_{\mbox{{\scriptsize eff}}}(\Delta B=-1)=\frac{G_{\mbox{{\scriptsize 
F}}}}{\sqrt{2}}\left[\sum\limits_{j=u,c}V_{jq}^\ast V_{jb}\left\{\sum
\limits_{k=1}^2Q_k^{jq}\,C_k(\mu)+\sum\limits_{k=3}^{10}Q_k^{q}\,C_k(\mu)
\right\}\right].
\end{equation}
Here $\mu={\cal O}(m_b)$, $Q_k^{jq}$ are four-quark operators, the label
$q\in\{d,s\}$ corresponds to $b\to d$ and $b\to s$ transitions, and $k$ 
distinguishes between current--current $(k\in\{1,2\})$, QCD 
$(k\in\{3,\ldots,6\})$ and EW $(k\in\{7,\ldots,10\})$ penguin operators. 
The evaluation of such low-energy effective Hamiltonians has been reviewed in 
Ref.\ 11, where the four-quark operators are given explicitly 
and numerical values for their Wilson coefficient functions can be found.

\subsection{CP Asymmetries in Decays of Neutral B Mesons}\label{CP-asym-neut}
A particularly simple and interesting situation arises, if we restrict 
ourselves to decays of neutral $B_q$ mesons ($q\in\{d,s\}$) into CP 
self-conjugate final states $|f\rangle$, satisfying the relation 
$({\cal CP})|f\rangle=\pm\,|f\rangle$. In this case, the corresponding 
time-dependent CP asymmetry can be expressed as
\begin{eqnarray}
\lefteqn{a_{\mbox{{\scriptsize CP}}}(t)\equiv\frac{\Gamma(B^0_q(t)\to f)-
\Gamma(\overline{B^0_q}(t)\to f)}{\Gamma(B^0_q(t)\to f)+
\Gamma(\overline{B^0_q}(t)\to f)}=}\nonumber\\
&&{\cal A}^{\mbox{{\scriptsize dir}}}_{\mbox{{\scriptsize CP}}}(B_q\to f)
\cos(\Delta M_q\,t)+{\cal A}^{\mbox{{\scriptsize
mix--ind}}}_{\mbox{{\scriptsize CP}}}(B_q\to f)\sin(\Delta M_q\,t)
\,,\label{ee6}
\end{eqnarray}
where the direct CP-violating contributions have been separated from
the mixing-induced CP-violating contributions, which are characterized by
\begin{equation}\label{ee7}
{\cal A}^{\mbox{{\scriptsize dir}}}_{\mbox{{\scriptsize CP}}}(B_q\to f)\equiv
\frac{1-\bigl|\xi_f^{(q)}\bigr|^2}{1+\bigl|\xi_f^{(q)}\bigr|^2}\quad
\mbox{and}\quad
{\cal A}^{\mbox{{\scriptsize mix--ind}}}_{\mbox{{\scriptsize
CP}}}(B_q\to f)\equiv\frac{2\,\mbox{Im}\,\xi^{(q)}_f}{1+\bigl|\xi^{(q)}_f
\bigr|^2}\,,
\end{equation}
respectively. Here direct CP violation refers to CP-violating effects
arising directly in the corresponding decay amplitudes, whereas 
mixing-induced CP violation is due to interference between 
$B_q^0$--$\overline{B_q^0}$ mixing and decay processes. Note that 
Eq.~(\ref{ee6}) has to be modified in the $B_s$ case for 
$t\mathrel{\hbox{\rlap{\hbox{\lower4pt\hbox{$\sim$}}}\hbox{$>$}}}1/
|\Delta\Gamma_s|$ because of the expected sizeable width difference 
$\Delta\Gamma_s$.\,\cite{dun}\, In general, the observable\,\cite{rev}
\begin{equation}\label{xi-expr}
\xi_f^{(q)}=\mp\,e^{-i\phi_{\mbox{{\scriptsize M}}}^{(q)}}\,
\frac{\sum\limits_{j=u,c}V_{jr}^\ast V_{jb}\bigl\langle f\bigl|{\cal Q}^{jr}
\bigr|\overline{B^0_q}\bigr\rangle}{\sum\limits_{j=u,c}V_{jr}
V_{jb}^\ast\bigl\langle f\bigl|{\cal Q}^{jr}
\bigr|\overline{B^0_q}\bigr\rangle}\,,
\end{equation}
where $\,{\cal Q}^{jr}\equiv\sum\limits_{k=1}^2Q_k^{jr}C_k(\mu)+
\sum\limits_{k=3}^{10}Q_k^{jr}C_k(\mu),\,$ and where 
\begin{equation}
\phi_{\mbox{{\scriptsize M}}}^{(q)}=\left\{\begin{array}{cr}
2\beta&\mbox{for $q=d$}\\
0&\mbox{for $q=s$}\end{array}\right. 
\end{equation}
denotes the weak $B_q^0$--$\overline{B_q^0}$ mixing phase, suffers from 
large uncertainties, which are introduced by the hadronic matrix elements 
in Eq.~(\ref{xi-expr}). There is, however, a very important special case. 
If the decay $B_q\to f$ is dominated by a single CKM amplitude, these
matrix elements cancel, and $\xi_f^{(q)}$ takes the simple form
\begin{equation}\label{ee10}
\xi_f^{(q)}=\mp\exp\left[-i\left(\phi_{\mbox{{\scriptsize M}}}^{(q)}-
\phi_{\mbox{{\scriptsize D}}}^{(f)}\right)
\right],
\end{equation}
where $\phi_{\mbox{{\scriptsize D}}}^{(f)}$ is a weak decay phase, which 
is given as follows ($r\in\{d,s\}$):
\begin{equation}\label{e11}
\phi_{\mbox{{\scriptsize D}}}^{(f)}=\left\{\begin{array}{cc}
-2\gamma&\mbox{for dominant $\bar b\to\bar u\,u\,\bar r$ CKM amplitudes
in $B_q\to f$}\\
0&\,\mbox{for dominant $\bar b\to\bar c\,c\,\bar r\,$ CKM amplitudes
in $B_q\to f$.}
\end{array}\right.
\end{equation}

Probably the most important applications of this well-known formalism are the 
decays $B_d\to J/\psi\, K_{\mbox{{\scriptsize S}}}$ and $B_d\to\pi^+\pi^-$.
If one goes through the relevant Feynman diagrams contributing to the former
channel (for a detailed discussion, see Ref.\ 10), one finds that it 
is dominated by the $\bar b\to\bar c\,c\,\bar s$ CKM amplitude. Consequently, 
the decay phase vanishes, and we have
\begin{equation}\label{e12}
{\cal A}^{\mbox{{\scriptsize mix--ind}}}_{\mbox{{\scriptsize
CP}}}(B_d\to J/\psi\, K_{\mbox{{\scriptsize S}}})=+\sin[-(2\beta-0)]\,.
\end{equation}
Since Eq.\ (\ref{ee10}) applies with excellent accuracy to the decay
$B_d\to J/\psi\, K_{\mbox{{\scriptsize S}}}$ -- the point is that penguins
enter essentially with the same weak phase as the leading tree
contribution -- it is usually referred to as the ``gold-plated'' mode
to determine the CKM angle $\beta$.\,\cite{csbs}\, First attempts to
measure $\sin(2\beta)$ through the CP-violating asymmetry (\ref{e12}) 
have recently been performed by the OPAL and CDF collaborations. Their
results are as follows:\,\cite{sin2b-exp}
\begin{equation}
\sin(2\beta)=\left\{\begin{array}{ll}
3.2^{+1.8}_{-2.0}\pm0.5&\mbox{OPAL Collaboration}\\
1.8\pm1.1\pm0.3&\,\mbox{\,CDF\, Collaboration;}
\end{array}\right.
\end{equation}
they favour the Standard Model expectation of a {\it positive} value of
this quantity. The presently allowed range arising from the usual fits
of the unitarity triangle is given by $0.36
\mathrel{\hbox{\rlap{\hbox{\lower4pt\hbox{$\sim$}}}\hbox{$<$}}}\sin(2\beta)
\mathrel{\hbox{\rlap{\hbox{\lower4pt\hbox{$\sim$}}}\hbox{$<$}}}
0.80$.\,\cite{burasHF}\, In the $B$-factory era, an experimental uncertainty 
of $\left.\Delta\sin(2\beta)\right|_{\rm exp}=0.08$ seems to be achievable.

In the case of the decay $B_d\to\pi^+\pi^-$, mixing-induced CP violation 
would measure $-\sin(2\alpha)$ through
\begin{equation}\label{e13}
{\cal A}^{\mbox{{\scriptsize mix--ind}}}_{\mbox{{\scriptsize
CP}}}(B_d\to\pi^+\pi^-)=-\sin[-(2\beta+2\gamma)]=-\sin(2\alpha)\,,
\end{equation}
if there were no penguin contributions present. However, we have to deal 
with such topologies, leading to hadronic corrections to Eq.~(\ref{e13})
that were analysed by many 
authors.\,\cite{alpha-uncert}$^{\mbox{-}}$\cite{charles}\, Last year, the 
CLEO collaboration reported the observation of several exclusive $B$-meson 
decays into two light pseudoscalar mesons.\,\cite{cleo}\, However, 
$B\to\pi\pi$ modes have not yet been seen and the upper limits for the 
corresponding branching ratios are not ``favourable''. The recent CLEO 
results indicate moreover that we have in fact to worry about the penguin 
corrections to Eq.\ (\ref{e13}). 

There are various methods on the market to control the penguin uncertainties 
in a quantitative way. Unfortunately, these strategies are usually rather 
challenging in practice. The best known approach was proposed by Gronau 
and London.\,\cite{gl}\, It makes use of the $SU(2)$ isospin relations
\begin{eqnarray}
\sqrt{2}\,A(B^+\to\pi^+\pi^0)&=&A(B^0_d\to\pi^+\pi^-)+
\sqrt{2}\,A(B^0_d\to\pi^0\pi^0)\label{iso1}\\
\sqrt{2}\,A(B^-\to\pi^-\pi^0)&=&A(\overline{B^0_d}\to\pi^+\pi^-)+
\sqrt{2}\,A(\overline{B^0_d}\to\pi^0\pi^0)\,,\label{iso2}
\end{eqnarray}
which can be represented in the complex plane as two triangles. The sides 
of these triangles are determined through the corresponding branching ratios, 
while their relative orientation can be fixed by measuring the CP-violating
observable ${\cal A}^{\mbox{{\scriptsize mix--ind}}}_{\mbox{{\scriptsize
CP}}}(B_d\to\pi^+\pi^-)$. Following these lines, it is in principle possible 
to extract a value of $\alpha$ taking into account the QCD penguin 
contributions. It should be noted that EW penguins cannot be controlled 
using this isospin strategy. Their effect is, however, expected to be 
rather small in this case, leading to $|\Delta\alpha_{\rm EW}/\sin\alpha|
\mathrel{\hbox{\rlap{\hbox{\lower4pt\hbox{$\sim$}}}\hbox{$<$}}}
4^\circ$.\,\cite{PAPIII}\, An attempt to analyse other isospin-breaking
effects, which may affect the isospin relations (\ref{iso1}) and (\ref{iso2})
through $\pi^0\,$--$\,\eta$,\,$\eta'$ mixing, was recently performed in
Ref.\ 22. Unfortunately, the Gronau--London approach suffers from an 
experimental problem, since the measurement of $B_d\to\pi^0\pi^0$ decays
is very difficult. Theoretical estimates based on the ``factorization'' 
hypothesis (for a critical look at this concept, see Ref.\ 23) give branching 
ratios at the $10^{-6}$ level.\,\cite{AKL1}\, However, upper bounds on the 
combined branching ratio BR$(B_d\to\pi^0\pi^0)$, i.e.\ averaged over the
decay and its charge conjugate, may already lead to interesting upper bounds 
on the QCD penguin uncertainty affecting the determination of 
$\alpha$.\,\cite{charles,gq-alpha} 

An alternative to the Gronau--London strategy to extract the CKM angle 
$\alpha$ is provided by $B\to\rho\,\pi$ modes.\,\cite{Brhopi}\, Here 
the isospin triangle relations (\ref{iso1}) and (\ref{iso2}) are replaced by 
pentagonal relations, and the corresponding approach is rather complicated.
The $SU(3)$ flavour symmetry offers also ways to determine $\alpha$. For
example, it is possible to extract this angle with the help of a triangle 
construction by measuring in addition to the $B_d\to\pi^+\pi^-$ observables 
those of $B_d\to K^0\overline{K^0}$ decays.\,\cite{PAPII}\, The latter 
would also be interesting to obtain insights into certain final-state 
interaction processes.\,\cite{bfm}\, A measurement of both the direct and  
the mixing-induced CP asymmetries in $B_d\to\pi^+\pi^-$, together with the 
$B^+\to\pi^+ K^0$ branching ratio, would provide another step towards the 
control of the QCD penguin uncertainties.\,\cite{fm1}\, Several strategies 
to constrain and determine $\alpha$ along these lines were recently proposed 
in Ref.\ 18. As sketched above, a solid extraction of this 
CKM angle is unfortunately quite difficult and could be out of reach for 
the first generation of $B$-factory experiments.

A decay appearing frequently in the literature as a tool to determine the 
CKM angle $\gamma$ is $B_s\to\rho^0 K_{\rm S}$. In this case, however, 
penguins are expected to lead to serious problems -- even more serious than 
in $B_d\to\pi^+\pi^-$ -- so that this mode appears to be the ``wrong'' way 
to extract $\gamma$.\,\cite{rev}\, Moreover, the rapid 
$B^0_s$--$\overline{B^0_s}$ oscillations, as well as the small expected 
branching ratio at the $10^{-7}$ level, make experimental studies of 
$B_s\to\rho^0 K_{\rm S}$ very difficult. It should be kept in mind, however, 
that this channel may be in better shape to probe $\gamma$, if the concept 
of ``colour suppression'' should not work in this case. A recent model 
calculation within a perturbative framework can be found in Ref.\ 30. 

\subsection{CP Violation in Penguin Modes as a Probe of New Physics}
In order to test the Standard Model description of CP violation, 
penguin-induced modes play an important role. Because of the loop suppression
of these ``rare'' processes, it is possible -- and indeed it is the case in
several specific model calculations -- that new-physics contributions to 
these decays are of similar magnitude as those of the 
Standard Model.\,\cite{new-phys}\, 

An important example is the $b\to s$ penguin mode $B_d\to\phi\, K_{\rm S}$. 
The corresponding branching ratio is expected to be of ${\cal O}(10^{-5})$ 
and may be large enough to investigate this channel at the future 
$B$-factories.\,\cite{AKL1}\, In contrast to the $b\to d$ penguin 
case,\,\cite{bff} the corresponding decay amplitude does not contain a 
sizeable CP-violating weak phase within the Standard Model. Consequently, 
direct CP violation in $B_d\to\phi\, K_{\rm S}$ is tiny, and mixing-induced 
CP violation measures simply the weak $B^0_d$--$\overline{B^0_d}$ mixing 
phase, implying the relation
\begin{equation}\label{SM-rel}
{\cal A}^{\mbox{{\scriptsize mix--ind}}}_{\mbox{{\scriptsize
CP}}}(B_d\to J/\psi\, K_{\mbox{{\scriptsize S}}})={\cal 
A}^{\mbox{{\scriptsize mix--ind}}}_{\mbox{{\scriptsize
CP}}}(B_d\to \phi\, K_{\mbox{{\scriptsize S}}})=-\sin(2\beta)\,,
\end{equation}
which represents an interesting probe of new-physics contributions to 
$b\to s$ decay processes. The theoretical accuracy of this relation is 
limited by certain neglected terms that are CKM-suppressed by 
${\cal O}(\lambda^2R_b)$, and may lead to direct CP violation
in $B_d\to\phi\, K_{\rm S}$. Simple model calculations performed at the 
perturbative quark level indicate asymmetries of at most 
${\cal O}(1\%)$.\,\cite{rev,AKL2}\, However, the impact of long-distance 
effects is hard to quantify (for a recent attempt, see Ref.\ 32). 
The importance of $B_d\to\phi\, K_{\rm S}$ and similar modes, such as 
$B_d\to\eta' K_{\rm S}$, to search for new-physics effects in $b\to s$ 
flavour-changing neutral current processes was emphasized by several 
authors.\,\cite{rev,AKL2,BdPhiKs} 

Studies of CP-violating effects in inclusive $B\to X_s\gamma$ decays, which
can be analysed reliably in QCD by means of the operator product 
expansion,\,\cite{kn-bsg} also play an important role in the search for new 
physics. Within the Standard Model, direct CP violation in $B\to X_s\gamma$ 
is very small, i.e.\ below the $1\%$ level, whereas it may well be as large 
as $50\%$ in new-physics scenarios with enhanced chromomagnetic dipole 
operators.\,\cite{kn}

\subsection{The $B_s$ System in the Light of $\Delta\Gamma_s$}\label{Bssys}
In the $B_s$ system, very rapid $B^0_s$--$\overline{B^0_s}$ oscillations are
expected, requiring an excellent vertex resolution system. Studies of CP 
violation in $B_s$ decays are therefore regarded as being very difficult. 
An alternative route to investigate CP-violating effects may be provided by 
the width difference $\Delta\Gamma_s$.\,\cite{DG-calc}\, Because of 
this width difference, already {\it untagged} data samples of $B_s$ decays 
may exhibit CP-violating effects.\,\cite{dun}\, Several ``untagged 
strategies'' to extract the CKM angle $\gamma$ were proposed, using for 
example angular distributions in $B_s\to K^{\ast+}K^{\ast-},\,K^{\ast0}
\overline{K^{\ast0}}$ or $B_s\to D^{\ast}\phi,\,D_s^{\ast\pm}K^{\ast\mp}$ 
decays.\,\cite{fd1}$^{\mbox{-}}$\cite{BpiKBsKK}\, 

The $B_s$ system provides interesting probes also for physics beyond the 
Standard Model. Important examples are the decays $B_s\to D_s^+D_s^-$ and 
$B_s\to J/\psi\,\phi$. The latter is the counterpart of the ``gold-plated'' 
mode $B_d\to J/\psi\,K_{\rm S}$ to measure $\beta$ and is very promising
for experiments performed at future hadron machines. These transitions are 
dominated by a single CKM amplitude and allow -- in principle even from 
their untagged data samples\,\cite{fd1} -- the extraction of a CP-violating 
weak phase $\phi_{\rm CKM}\equiv2\lambda^2\eta$, which is expected to be of 
${\cal O}(0.03)$ within the Standard Model. Consequently, an extracted 
value of $\phi_{\rm CKM}$ that is much larger than this 
Standard Model expectation would signal new-physics contributions to
$B^0_s$--$\overline{B^0_s}$ mixing.\,\cite{new-phys}\, Suggestions for 
efficient determinations of the observables of the $B_s\to D_s^{\ast+}
D_s^{\ast-}$ and $B_s\to J/\psi\,\phi$ angular distributions, as well as 
of $\Delta\Gamma_s$ and of the $B_s$ mass difference $\Delta M_s$, 
were given in Ref.\ 40.

A time-dependent study of $B_s\to J/\psi\,\phi$ decays is also interesting
to resolve a discrete ambiguity in the determination of the CKM angle 
$\beta$.\,\cite{ddf2}\, Such ambiguities are a typical feature
of the strategies to extract CKM phases sketched above; they could complicate 
the search for new physics considerably. Several strategies were recently
proposed to deal with these problems.\,\cite{ambig}

\subsection{CP Asymmetries in Decays of Charged B Mesons}
Since mixing effects are not present in the charged $B$-meson system, the 
measurement of a non-vanishing CP asymmetry in a charged $B$ decay would 
give us unambiguous evidence for direct CP violation, thereby ruling 
out ``superweak'' models.\,\cite{superweak}\, Such CP asymmetries arise from 
interference between decay amplitudes with both different CP-violating weak 
and CP-conserving strong phases. Whereas the weak phases are related to the 
CKM matrix, the strong phases are induced by strong final-state interaction 
effects and introduce in general severe theoretical uncertainties into the 
calculation. An interesting situation arises, however, in 
$B^\pm\to\rho^\pm\rho^0(\omega)\to\rho^\pm\pi^+\pi^-$ decays, where 
$\rho^0(\omega)$ denotes the $\rho^0\,$--$\,\omega$ interference 
region.\,\cite{et-got}\, In this case, experimental data on 
$e^+e^-\to\pi^+\pi^-$ processes can be used to constrain the hadronic 
uncertainties affecting the corresponding CP asymmetry, which is related  
to $\sin\alpha$ and may well be as large as ${\cal O}(20\%)$ at the $\omega$ 
invariant mass. Direct CP violation in three-body decays such 
as $B^\pm\to K^\pm\pi^+\pi^-$, which involve various intermediate resonances, 
was recently considered in Ref.\ 44. Here the Dalitz plot distributions 
may provide information on the CKM angle $\gamma$ (see also Ref.\ 45). 

In order to extract angles of the unitarity triangle from decays of charged 
$B$ mesons, amplitude relations -- either exact or approximate ones 
based on flavour symmetries --  play an important role. A review of these 
methods can be found in Ref.\ 10. The ``prototype'' is the approach 
to determine $\gamma$ with the help of triangle relations between the 
$B^\pm\to D K^\pm$ decay amplitudes proposed by Gronau and Wyler.\,\cite{gw}\, 
Unfortunately, the corresponding triangles are expected to be very 
``squashed''. Moreover, one has to deal with additional experimental 
problems,\,\cite{ads} so that this approach is very difficult from a 
practical point of view. More refined variants and different ways to
combine the information provided by $B\to DK$ decays to probe the CKM angle
$\gamma$ were proposed by several authors.\,\cite{ads,bdk}\, The CLEO
collaboration has recently observed the colour-allowed decay $B^-\to D^0K^-$
and its charge conjugate, which is the first detection of a $B$ decay
originating from $b\to c\,\bar u\,s$ quark subprocesses.\,\cite{cleo-bdk}

\subsection{Strategies to Probe the CKM Angle $\gamma$ with $B\to\pi K$ Modes}
The decays $B^+\to\pi^+K^0$, $B^0_d\to\pi^-K^+$ and their charge conjugates,
which were observed by the CLEO collaboration last year,\,\cite{cleo} 
play an important role to probe the CKM angle $\gamma$ at future 
$B$-factories.\,\cite{PAPIII,groro}\, So far, only results for the combined 
branching ratios BR$(B^\pm\to\pi^\pm K)$ and BR$(B_d\to\pi^\mp K^\pm)$
have been published at the $10^{-5}$ level with large experimental 
uncertainties. In order to obtain information on $\gamma$, the ratio
\begin{equation}\label{Def-R}
R\equiv\frac{\mbox{BR}(B_d\to\pi^\mp K^\pm)}{\mbox{BR}(B^\pm\to\pi^\pm K)}
\end{equation}
of the combined $B\to\pi K$ branching ratios, and the ``pseudo-asymmetry'' 
\begin{equation}
A_0\equiv\frac{\mbox{BR}(B^0_d\to\pi^-K^+)-\mbox{BR}(\overline{B^0_d}\to
\pi^+K^-)}{\mbox{BR}(B^+\to\pi^+K^0)+\mbox{BR}(B^-\to\pi^-\overline{K^0})}
\end{equation}
play a key role. Making use of the $SU(2)$ isospin symmetry of strong
interactions, the following amplitude relations can be derived (for a
detailed discussion, see Ref.\ 28):
\begin{equation}
A(B^+\to\pi^+K^0)\equiv P\,,\quad A(B_d^0\to\pi^-K^+)=-\,\left[P+T+P_{\rm ew}
\right],
\end{equation}
where the ``penguin'' amplitude $P$ is {\it defined} by the $B^+\to\pi^+K^0$ 
decay amplitude, $P_{\rm ew}\equiv-\,|P_{\rm ew}|e^{i\delta_{\rm ew}}$ is 
essentially due to electroweak penguins, and 
$T\equiv|T|e^{i\delta_T}e^{i\gamma}$ is usually referred to as a ``tree'' 
amplitude. However, because of a subtlety in the implementation of the 
isospin symmetry, the amplitude $T$ does not only receive contributions from 
colour-allowed tree-diagram-like topologies, but also from penguin and 
annihilation topologies.\,\cite{bfm,defan}\, The general expressions for 
the amplitudes $P$, $T$ and $P_{\rm ew}$ are given in Ref. 51. Let us just 
note here that we have
\begin{equation}
P\propto\left[1+\rho\,e^{i\theta}e^{i\gamma}
\right],
\end{equation}
where $\rho\,e^{i\theta}$ is a measure of the strength of certain 
rescattering effects ($\theta$ is a CP-conserving strong phase). The 
quantities $r\equiv|T|/\sqrt{\langle|P|^2\rangle}$ and $\epsilon\equiv
|P_{\rm ew}|/\sqrt{\langle|P|^2\rangle}$ with 
$\langle|P|^2\rangle\equiv(|P|^2+|\overline{P}|^2)/2$, as well as the
strong phase differences $\delta\equiv\delta_T-\delta_{tc}$ and $\Delta
\equiv\delta_{\rm ew}-\delta_{tc}$, where $\delta_{tc}$ measures the
strong phase of the difference of the penguin topologies with internal top 
and charm quarks, turn out to be very useful to parametrize the observables 
$R$ and $A_0$.\,\cite{defan}\,

If both $R$ and $A_0$ are measured, contours in the $\gamma\,$--$\,r$ plane 
can be fixed,\,\cite{defan} corresponding to a mathematical implementation 
of a simple triangle construction.\,\cite{PAPIII}\, In order to determine 
the CKM angle $\gamma$, the quantity $r$, i.e.\ the magnitude of the ``tree'' 
amplitude $T$, has to be fixed. At this step, a certain model dependence 
enters. Using arguments based on ``factorization'', one comes to the 
conclusion that a future theoretical uncertainty of $r$ as small as 
${\cal O}(10\%)$ may be achievable.\,\cite{groro,wuegai}\, However, 
since the properly defined amplitude $T$ does not only receive 
contributions from colour-allowed ``tree'' topologies, but also from penguin 
and annihilation processes,\,\cite{bfm,defan} it may be shifted sizeably 
from its ``factorized'' value so that $\Delta r={\cal O}(10\%)$ may be 
too optimistic. 

Interestingly, it is possible to derive bounds on $\gamma$ that do {\it not}
depend on $r$ at all\,\,\cite{fm2} (for a discussion of $B_s\to K\overline{K}$
decays, see Ref.\ 39). If we eliminate the strong phase $\delta$ 
in the ratio $R$ of combined $B\to\pi K$ branching ratios with the help of
the pseudo-asymmetry $A_0$ and treat $r$ as a ``free'' variable, while keeping
$\rho$, $\theta$ and $\epsilon$, $\Delta$ fixed, we find that $R$ takes the 
following minimal value:\,\cite{defan} 
\begin{equation}\label{Rmin}
R_{\rm min}=\kappa\,\sin^2\gamma\,+\,
\frac{1}{\kappa}\left(\frac{A_0}{2\,\sin\gamma}\right)^2.
\end{equation}
In this expression, rescattering and EW penguin effects are described by 
\begin{equation}
\kappa=\frac{1}{w^2}\left[\,1+2\,(\epsilon\,w)\,\cos\Delta\,+\,
(\epsilon\,w)^2\,\right]\quad\mbox{with}\quad 
w=\sqrt{1+2\,\rho\,\cos\theta\,\cos\gamma+\rho^2}. 
\end{equation}
An allowed range for $\gamma$ is related to $R_{\rm min}$, since values of
$\gamma$ implying $R_{\rm exp}<R_{\rm min}$, where $R_{\rm exp}$ denotes
the experimentally determined value of $R$, are excluded. In the ``original'' 
bounds on $\gamma$ derived in Ref.\ 53, no information provided by $A_0$ 
has been used, i.e.\ both $r$ and $\delta$ were kept as ``free'' variables, 
and the special case $\rho=\epsilon=0$, i.e.\ $\kappa=1$, has been assumed, 
implying $\sin^2\gamma<R_{\rm exp}$. A particularly interesting situation 
arises, if $R$ is measured to be smaller than~1. In this case, the information
on $\gamma$ provided by the $B\to\pi K$ decays is complementary to the 
presently allowed range of 
$41^\circ\mathrel{\hbox{\rlap{\hbox{\lower4pt\hbox{$\sim$}}}\hbox{$<$}}}\gamma
\mathrel{\hbox{\rlap{\hbox{\lower4pt\hbox{$\sim$}}}\hbox{$<$}}}134^\circ$, 
which arises from the usual fits of the unitarity triangle,\,\cite{burasHF}
since a certain interval around $90^\circ$ can be excluded.\,\cite{fm2}\, 
A measurement of $A_0\not=0$ allows us to exclude in addition a range around 
$0^\circ$ and $180^\circ$.\,\cite{defan}\, Unfortunately, the 
present data do not yet provide an answer to the question of whether $R<1$. 
The results reported by the CLEO collaboration last year give 
$R=0.65\pm0.40$,\,\cite{cleo} whereas a very recent, updated analysis yields 
$R=1.0\pm0.4$.\,\cite{jimA}

The theoretical accuracy of these bounds is limited both by rescattering 
processes of the kind $B^+\to\{\pi^0K^+,\,\pi^0K^{\ast +},\,\ldots\}\to
\pi^+K^0$,\,\cite{FSI}$^{\mbox{-}}$\cite{fknp} and by EW penguin 
effects.\,\cite{groro,neubert}\, An important implication of the
rescattering effects may be a sizeable CP asymmetry, $A_+$, in the
decay $B^+\to\pi^+K^0$. The rescattering effects can be controlled in the 
bounds on $\gamma$ through experimental data. A first step towards this goal 
is provided by the CP asymmetry $A_+$. In order to go beyond it, 
$B^\pm\to K^\pm K$ decays -- the $SU(3)$ counterparts of 
$B^\pm\to \pi^\pm K$ -- play a key role,\,\cite{defan,rf-FSI} allowing us to 
include the rescattering processes in the contours in the $\gamma\,$--$\,r$ 
plane and the associated constraints on $\gamma$ completely  
(for alternative strategies, see Refs.\ 28 and 57). Since 
the ``short-distance'' expectation for the combined branching ratio
BR$(B^\pm\to K^\pm K)$ is ${\cal O}(10^{-6})$,\,\cite{AKL1} experimental 
studies of $B^\pm\to K^\pm K$ appear to be difficult. These modes have not 
yet been observed, and only upper limits for BR$(B^\pm\to K^\pm K)$ are 
available.\,\cite{cleo,jimA}\, However, rescattering effects may enhance this 
quantity significantly, and could thereby make $B^\pm\to K^\pm K$ measurable 
at future $B$-factories.\,\cite{defan,rf-FSI}\, Another important indicator 
of large FSI effects is provided by $B_d\to K^+K^-$ decays,\,\cite{groro-FSI} 
for which stronger experimental bounds already exist.\,\cite{cleo,jimA}\,

In the case of the decays $B^+\to\pi^+K^0$ and $B^0_d\to\pi^-K^+$, EW penguins
contribute only through ``colour-suppressed'' topologies; estimates based 
on the ``factorization'' hypothesis typically give values for $\epsilon$ 
at the $1\%$ level.\,\cite{AKL1}\, These crude estimates may, however, 
underestimate the role of the EW penguins.\,\cite{groro,neubert}\, An 
improved description is possible by using the general expressions for the 
EW penguin operators and performing appropriate Fierz transformations; on
the way to controlling the corresponding uncertainties in the bounds on 
$\gamma$ with the help of experimental data, the decay $B^+\to\pi^+\pi^0$ 
provides an important first step.\,\cite{defan}\,

\section{Conclusions and Outlook}
In conclusion, we have seen that certain non-leptonic $B$ decays offer a 
fertile ground to test the Standard Model description of CP violation.
Detailed experimental studies of these modes at $B$-factories are just 
ahead of us and may bring unexpected results, which could guide us to the 
physics lying beyond the Standard Model. The coming years will certainly 
be very exciting!

\section*{Acknowledgements}
I would like to thank the organizers, in particular Professors Hai-Yang
Cheng, Guey-Lin Lin and Hoi-Lai Yu, for inviting me to this stimulating 
workshop, for their kind hospitality, and for covering my local expenses. 
The travel support from the {\it Deutsche Forschungsgemeinschaft} is also 
gratefully acknowledged.

\vspace*{0.6cm}
\par

\end{document}